\documentclass{article}
\usepackage{latexsym}
\usepackage{amsmath}
\usepackage{amsfonts}
\usepackage{amssymb}
\newcommand{\BC}{\mathbb{C}}

\newcommand{\BP}{\mathbb{P}}
\newcommand{\beq}{\begin{eqnarray}}
\newcommand{\eeq}{\end{eqnarray}}
\newcommand{\five}[1]{\overset{5}{#1}}
 
\begin{document}
\title{Projective Relativity: Present Status and Outlook\thanks{%
This paper is dedicated to Prof. Dehnen on the occasion of his 65th
anniversary.}}
\author{Bertfried Fauser\\
Universit\"at Konstanz\\
Fachbereich Physik, Fach M678\\
78457 Konstanz, Germany\\
E-mail: Bertfried.Fauser@uni-konstanz.de
}
\date{October 29, 2000}
\maketitle

\begin{abstract}
We give a critical analysis of projective relativity theory. Examining
Kaluza's own intention and the following development by Klein, Jordan,
Pauli, Thiry, Ludwig and others, we conclude that projective relativity 
was abused in its own terms. Much more in the case of newer higher 
dimensional Kaluza--Klein theories with non-Abelian gauge groups. 
Reviewing the projective formulation of the Jordan isomorphy theorem 
yields some hints how one can proceed in a different direction. We can 
interpret the condition $\five{R}_{\mu\nu}=0$ not as a field equation in a 
5-dimensional Riemannian space, e.g. as vacuum Einstein-Hilbert 
equation, but can (or should) interpret it as a geometrical object, a 
null-quadric. Projective aspects of quantum (field) theory are discussed 
under this viewpoint. 


\end{abstract}

\section{Introduction}

\subsection{Our Motivation}

The aim of the present note is to oppose projective geometry to such 
different branches of physics as projective relativity and quantum 
field theory. We unmask a hidden Riemannian core of projective 
relativity, the origin of the vacuum field equations, in a polemic way
using Hestenes' idea of `mathematical viruses'. We belief, that only a
radical projective point is able to reformulate relativity theory
in such a way, that compatibility to quantum field theory can be 
obtained. In order to to this, we propose the null-quadric to be
taken a `principle' and hence as origin of the 5-dimensional field 
equations. This argument is valid in 5 dimensional Kaluza-Klein 
theories only. Projective differential geometry was surprizingly already 
developed very early \cite{Wil,Hessenberg} even if used marginally later.

Trying to compactify and understand the algebraic structure of quantum
field theory, we have been led in a series of papers 
\cite{Fauser-positron,Fauser-dirac,Fauser-vertex,Fauser-thesis,%
Fauser-vacua,Fauser-transition,Fauser-wick} to two main results:
Gra{\ss}mann Hopf and Clifford (Hopf?) algebras are the natural language
to code quantum field theories. Some singularities simply drop out
after a rigorous algebraization; and, Clifford geometry and
Gra{\ss}mann-Cayley projective structure are not only related to but part
of the core features of quantum field theory.

We try to puzzle together projective relativity and projective quantum 
field theory to contrast recent belief in quantization of gravity e.g. 
by super string, brane or M-theory. However, our arguments are not yet 
fully developed and a deeper understanding of the Hopf algebraic structure 
involved in both cases is needed, but first outcomes are summarized in 
the conclusion.

\subsection{Kaluza's Motivation}

In 1921 a paper by Th. Kaluza on the unity problem of physics 
\cite{Kaluza} found its way into print only after its publication had been
postponed several times by Einstein who seemed to have disliked this idea.
Kaluza, making a recourse to Weyl, tries to interpret the electromagnetic
four potential $A_\mu$ and the therefrom derived field strength $F_{\mu\nu}$
as a `verst\"ummelte Dreizeigergr\"o{\ss}e' [truncated Christoffel symbol].
Comparing $1/2F_{\mu\nu}=1/2(A_{\mu,\nu}-A_{\nu,\mu})$ with 
$[\,{{i\lambda}\atop{\kappa}}\,]=\,1/2(g_{i\kappa,\lambda}+
g_{\kappa\lambda,i}-g_{i\lambda,\kappa})$ he concluded that one can 
unite gravity and electromagnetism by the introduction of a fifth world
dimension. This leads to the new 5-dimensional `metric'
\begin{equation}
(\five{g}_{ij}) := \left(\begin{array}{cc}
g_{\mu\nu} & A_{\kappa} \\
A^T_{\lambda} & \phi
\end{array}\right).
\end{equation}
Kaluza immediately runs into the problem, that physically observed 
quantities do not depend on the fifth world coordinate. He circumvents
this problem by introducing the `cylinder condition' i.e. the requirement 
that $\partial_5 (\ldots) = 0$ for any physical quantity. A second problem
is the scalar $\five{g}_{55}=\phi$ which had not an immediate interpretation
and was set to one.
However, after this artifical steps, Kaluza was able to incorporate the
electromagnetism in his 5-dimensional formalism in a beautiful and
satisfying way by splitting all equations into 1-4 dimensions and the 5th 
one. However, for our point the last sentence in Kaluza's paper is of great 
importance:
``Sollte es sich aber einmal best\"atigen, da{\ss} mehr hinter den
vermuteten Zusammenh\"angen steckt als nur ein leerer Formalismus, so
w\"urde dies entschieden einen neuen Triumph f\"ur Einsteins allgemeine
Relativit\"atstheorie bedeuten, um deren sinngem\"a{\ss}e Anwendung 
auf eine f\"unfdimensionale Welt es sich hier handelt'' \cite{Kaluza}. 
[Should more than a presumed formalism be found to reside behind these 
presumed connections, we would then face a new triumph of Einstein's general 
relativity, whose appropriate application to a five-dimensional world 
is our main concern here. \cite{Kaluza-eng}] Beside the probable wish to 
please Einstein, this sentence shows, that Kaluza was {\it not\/} led to 
his idea by projective reasoning, but by applying once more metrical 
structures in the additional 5th dimension. This became much more true in 
modern non-Abelian extensions of Kaluza--Klein theories.

\subsection{Projective Interpretation}
 
The beauty and compactness of Kaluza's unification attracted many 
other scientists to work in this area. Among others, Klein, Veblen,
Einstein(!), Jordan, Pauli, Ludwig, Thiry, Schmutzer, Lessner, 
\cite{Jordan,Pauli1,Pauli2,Ludwig,Schmutzer,Lessner} 
etc. and of course Dehnen \cite{MaciasDehnen}
developed a deeper understanding of Kaluza--Klein theories some of them 
by introducing projective techniques. 

In a metrical model of a projective $n$-space one can describe
projective `points' by equivalence classes of $n+1$ tuples of 
coordinates up to a scalar factor. A projective 4-space has thus 
5 coordinates. An analogous situation occurs in the reduction from 
4-dimensional Minkowski space to 3-dimensional Euclidean space of 
Newtonian physics. An obvious benefit of such an interpretation is, 
that one has not to bother about any metrical interpretation 
of the 5-dimensional space in terms of `world dimensions'. 

The main problem is to relate the projective space to a metrical
world continuum in an appropriate way. This was done by Jordan, 
proving the Jordan isomorphy theorem, which provides a reduction from
projective space to the usual Riemannian 4-space. It was also Jordan, 
who noticed that the new scalar $\phi$ introduced in the 5-metric
can be related to a field and the gravitational `constant'. This provides
a possibility to vary the strength of gravitation. Such a reasoning opened 
the way for developments as inflation and dynamical cosmological 
models.

However, the germ of a metrical theory could not be abandoned even in the
projective approaches developed so far, since there is a need to come up
with field equations in the 5-dimensional space and these field equations
are usually modelled in a way close to the Einstein-Hilbert theory
i.e. based on a metrical continuum. This will be discussed below in more
detail. We can conclude, that the projective interpretation was not 
fully adopted in Kaluza--Klein theories, which have been seen and treated 
as geo{\it metrical\/} theories all the times.

\section{Present Status --}

\subsection{Mathematical and Physical Viruses}

D. Hestenes introduced in \cite{Hes-virus} the notion of a mathematical 
virus (MV), which is an analogy of a biological virus (BV) or a computer 
virus (CV). ``A virus cannot exist by itself, but when attached to a host
it replicates repeatedly until it impairs the function of the host, 
sometimes to a point of disabling the host altogether'' \cite{Hes-virus} 
p.3. From this definition, it is clear that also ideas (or better ideologies)
can be treated as viruses. Hestenes proceeds to describe as species of 
mathematical viruses the coordinate virus MV/C: `Coordinates are essential 
to calculations', the quadratic form virus MV/Q: `Clifford algebra is the 
algebra of a quadratic form' which will become important below. Different 
other species of MV as MV/G, MV/DS, MV/K and MV/T are also described there,
and the reader is called to `infect' himself by reading the article.

We add to this list, as Varags has done recently \cite{Var-virus}
in an analogous context, two more mathematical or might be physical 
viruses (PV). The first is the point space virus:\\
{\bf PV/P: `Lines, planes, space, etc. are made from continua of points'}.\\ 
The antidote to this kind of virus is projective geometry, 
where one can in principle start from say planes to construct lines and 
points showing that points can be seen as a continuum of planes. 
The {\it principle of duality\/} (in projective 3 space) is simply the 
statement that if one interchanges `point' with `plane' and the operations 
`join' with `meet' and vice versa, every theorem for `points' will hold 
true for `planes' and vice versa. The theorem of Brianchon, a dual version
of Pascal's theorem is a prominent case, where it took 150 years to find
a truth which is a trivial outcome of this principle \cite{KadisonKromann}. 

Now, Riemannian geometry is intrinsically tied to the primacy of points,
since the metric is given infinitesimally at a point on a Riemannian 
manifold e.g.
\begin{eqnarray}
\mbox{d}s^2 &=& \five{g}_{ij}{\rm d}x^i{\rm d}x^j
\,=\, g_{\mu\nu}{\rm d}x^\mu{\rm d}x^\nu
+g_{5\mu}{\rm d}x^5{\rm d}x^\mu +g_{\mu 5}{\rm d}x^\mu{\rm d}x^5
+g_{55} {\rm d}x^5{\rm d}x^5.
\end{eqnarray}
But it is equally natural to ask for such an equation attached
to `planes'. Because of the fact that point and plane coordinates are
mutually dual one would probably recover similar equations. For
an projective approach to duality and Classical Mechanics see 
\cite{Con-B,Con-M}. A methodological approach to a projective formulation 
of physics from non-metrical grounds is given in 
\cite{Gschwind-RZG,Gschwind-QM}. The
most interesting point, however, is to ask for a self dual description.
In projective 3-space (this will be important later on) this leads naturally 
to `lines', since lines are dual to lines. It is possible to develop 
projective theory from lines. Even projective differential geometry 
{\it should\/} be developed {\it before\/} the advent of the metrical 
theories. Indeed, one can find an emphatical such statement in the book 
of E. Wilczynski \cite{Wil}, dated 1906, so protected from an infection by
the Einstein--Hilbert metrical theory, a virus? Moreover, one should not 
forget, that the natural space to investigate differential equations is 
projective space, as developed by Lie and Engel \cite{LieEng}. Point and 
contact transformations have there their origin.

The second virus was already discussed above, and could be called the 
manifold Virus:\\
{\bf PV/M: `Physics takes place on a Riemannian Manifold'}.\\ 
This was the belief of Kaluza, see the citation above. We have obviously 
non-Riemannian manifolds currently employed in physics. Fiber-bundles with 
non-Abelian gauge groups, super-spaces, non-commutative geometries, 
quantum field theories using strings, branes or M-theory, but all of them 
are `point' theories developed in this spirit, especially local quantum 
field theory, but see below.

\subsection{Projective Quantum Theory}

We can give only cursory arguments, some more (and orthogonal) arguments
can be found in \cite{Gschwind-QM}.

In e.g.\cite{Cartan-grpproj,Cartan-duality} and his book
\cite{Cartan-thofspin} Cartan developed the theory of projective spaces and
spinors. It is worth to note, that Cartan used spinors for algebraic
and geometric reasons in a time when they had not even be known in
physics. He showed furthermore, that one needs spinors to find all
irreducible representations of the classical groups. So, spinors are the
most natural and useful tool to study null-invariance and quadrics in
projective spaces. Null-invariance is a concept of the space of linear 
complexes in line geometry \cite{Stosz1,Stosz2}.

Already in \cite{BirkNeu} Birkhoff and von Neumann showed that quantum
mechanics is based on projective concepts. One has to note that the
`scalar product' of quantum mechanics is in fact a dual product, or
pairing between space and dual space, which contains no metrical 
information:
\beq
<\,\mid\,> : {\cal H}^\ast \times {\cal H} & \longrightarrow & \BC,
\eeq
where ${\cal H}^\ast$ is the dual space. One should note that the late
Dirac told, that projective reasonings led him to find his electron
equation in 1928 \cite{Rechenberg}. Dirac was furthermore interested in a
projective formulation of Hamilton's principle \cite{Dirac-HV} and the
usage of homogenous variables in classical physics. Indeed, it is possible
to find a dual and projective formulation of classical mechanics and
symplectic geometries \cite{Con-thesis}. Projective considerations seems to
have played a major role in electron physics, a fact which is not well
recognized because Hilbert space methods cannot clearly point this out

\paragraph{Quantum physics is incidence physics :}

If one analyses the mathematics of the quantum mechanical apparatus
\cite{Verhulst}, it is clear that in principle a system can only be prepared
at a time and measured at a second time. No information at a third time is
available in principle. The time evolution has to be a free undisturbed
evolution of the given dynamics described by a suitable Hamiltonian. If we
denote the events of preparation and measurements as `points', their
unknown connection can be addressed as a `line'. Such a line might in no
way be straight. Paths in Bohmian mechanics or streamlines of the
Dirac fluid show quite complicated patters. However, the important fact is
that by construction the Hamiltonian has to be a linear operator and
generates the time development. Moreover, to fix a `higher order' (i.e
non-geodesic) curve would necessarily involve at least a third `point' 
to be known. In this sense a virtual connection between preparation 
and measurement can be addressed as a `line'. Looking at the geodesic on 
a sphere, which are great circles, one observes that they are linearly 
parametrized by the angle. Only if one projects these `lines' by a 
non-linear, say stereographic, projection on a plane `lines' become 
`curves' in this plane w.r.t. the distance measure there. The mere notation
\beq
\psi(x,t)&:=&e^{i\hat{{\cal H}}t}\, \psi(x,0)
\eeq
is essentially of the same type.

\subsection{Algebraization of QFT}

A little bit surprising might be the fact that the method of `second
quantization' exhibits a projective structure. While the fermionic
creation operators $a_i^\dagger$ span a linear space $V$ their duals, the
annihilation operators, are elements of the dual space $V^\ast$. The Fock
`inner product' is thus once more a dual product
\beq
<\,\mid\,> &:& \bigwedge(V^\ast) \times \bigwedge(V) \longrightarrow \BC
\\ \nonumber
&& <n\mid m> = <0\mid a_{i_1}\ldots a_{i_n}
a_{j_1}^\dagger\ldots a_{j_m}^\dagger\mid 0> = \delta_{n,m}
\eeq
in an obvious notation. The advantage of such a point of view is, that it
allows an easy formulation of QFT in terms of the associated Clifford
algebras (symplectic Clifford algebras or Weyl algebras for
bosons) \cite{Fauser-thesis,Fauser-transition}. Such a formulation is
equivalent to `functional quantum field theory' as developed and
successfully utilized in \cite{Stumpf}. Resisting MV/Q, we note that 
Clifford algebras of dual-products do not depend on a metrical structure.

Much more surprising might be that a thoughtful formulation of QFT in
Clifford terms avoids a common singularity of QFTs due to the transition
from one ordering to another, see 
\cite{Fauser-vertex,Fauser-transition,Fauser-wick}. 
It seems to be a general rule that those concepts of QFT which are
geometrizable are `sane' while non-geometrizable operations, as e.g.
the Bethe-Salpeter equation, are `insane'. But, the large amount of
geometrical theorems leads to unsolved questions in QFT when transfered 
in this direction.

\subsection{Geometric Algebras}

A main problem in dealing with geometrical properties is the 
mathematical axiomatization of these notions. Indeed, projective geometry
and algebra are brother and sister originating from the same problems.
For our purpose, we mention Gra{\ss}mann \cite{GrassmannA1} and Clifford
\cite{Clifford}, while M\"obius, Pl\"ucker, Poncelet, Lindemann, Cayley 
and Klein, to cite only few, played also an important part. The main point 
for our argumentation is, that and algebraic description was developed which
allowed to manipulate symbolically and without the Cartesian coordinates
algebraic objects. The birth of algebraic geometry.

\paragraph{Clifford Geometric Algebras :}

Unfortunately projective geometry became old-fashioned and doomed away. Only 
a few persons, among them D. Hestenes see e.g. 
\cite{Hes-STA,Hes-linalg,HesZie},
had a continuous interest in `Geometric Algebra'. This was the term Clifford
used for Clifford algebras. However, there is not a single geometric algebra,
but algebraic structures are dependent on the semantic context and can be 
used to model geometric situations only after algebraic terms have been tied
to geometric meanings \cite{FauserAbla}. 

Nevertheless, we find lots of metrical applications of Clifford 
geometrical algebra e.g. in Lie group theory \cite{DoranLieGrps}. 
But Gra{\ss}mann developed his system in a projective background, as 
Clifford did also. 

Important improvements have been achieved in a series of papers 
which develop a Clifford geometric formulation of quantum field theories
\cite{Fauser-vertex,Fauser-thesis,Fauser-transition}.
The main tool in such a framework is the so-called {\it quantum Clifford
algebra\/}, already studied in 
\cite{Oziewicz-FGTC,Oziewicz97,Fauser-mandel,AblaLounesto}
and properly defined in \cite{FauserAbla}. However, all these treatments 
of quantum Clifford algebras share the methodological problem that the 
algebras are constructed starting with vector variables and seem to 
imply a metrical interpretation. This can be noticed in the the frequent 
statements --preconception or virus-- that: 
``Clifford algebras are metrical algebras'', a variant of MV/Q!

\paragraph{Gra{\ss}mann--Cayley and Hopf Algebras :}

A quite different approach starts with incidence relations which can
be coded in the lattice of subspaces of a linear space, see
\cite{Barnabei,Doubilet,Joni}. One defines algebraically two
operations called `join' and `meet' describing the linear span of 
two subspaces, two points span a line (join) etc. or their intersection, 
two  lines may or may not intersect (meet) in a point. Putting some rules 
how to calculate, one obtains a Gra{\ss}man-Cayley (GC) algebra 
\cite{Barnabei}. This algebra has also non geometrical applications in 
the theory of determinants, combinatorics etc., or the other way around, 
this theory provides a geometrization of large parts of mathematics. 
It is of recent interest, that the meet in GC algebras can be calculated 
in an straight forward manner using Hopf algebra techniques. The meet
(join) can be seen to be related to a co-product on the dual space.

It was Gian-Carlo Rota, who developed the theory of Peano space 
\cite{Barnabei} and GC algebra utilizing Hopf algebras. 
The co-product of a Hopf algebra is needed to compute in a clean way the
meet, which cannot be done using Gra{\ss}mann algebra alone. This is a
long ongoing and still present story starting with Gra{\ss}mann's 
regressive product, which lowers the grade. The most remarkable fact is, 
that by Cliffordization \cite{RotaStein} one can turn a GC algebra into 
a Clifford algebra. This process is intimately related to the 
normal-ordering process of quantum field theory \cite{Fauser-wick}. 
Moreover, using the Hopf algebra antipode, Connes and Kreimer 
\cite{Kreimer} have been able to produce the counter terms of
perturbative renormalization. For short, Feynman diagrams may be 
understood as tangles of certain Clifford or Gra{\ss}mann Hopf 
algebras: annihilation is the `product', creation is the `co-product'
and physics is done by convolution.

All this is a complicated not yet fully understood relation, but it
emerges directly from projective geometry which is recovered in the 
structure of quantum fields. Some of this observations belong already 
to the next part. 

\section{ -- and Outlook}

\paragraph{Jordan isomorphism theorem versus In\"on\"u-Wigner group
contractions :}

One of the basic problems of a 5 or higher dimensional Kaluza-Klein theory
is the way back to the observed $3+1$ dimensions. The main point of such a
development is an mechanism which breaks a symmetry into a semi-direct
product of sub-symmetries. A common example is the
non-relativistic limit $c\rightarrow \infty$ of the Lorentz group, a
classical group, into the not semi-simple Galileo group. Jordan, in
reducing the 5 dimensional Kaluza-Klein theory splited off the scale
transformations which have no effect on homogeneous coordinates, see
\cite{Ludwig}. Since a projective point is defined in a metrical model by
an $n+1$ tuple of coordinates $\BP^n \ni x=(x_1,\ldots,x_{n+1})$ modulo a
constant $\rho$ and not all $x_i=0$. One finds $\rho(x_1,\ldots,
x_{n+1})=(x_1,\ldots, x_{n+1})$ mod $\rho$ for such points. It is
therefore necessary to extract the transformations of the $(x_1,\ldots,
x_{n+1})$ which do not respect this properties. The relevant groups are
$GL(n+1)$ reducing to $PGL(n)$. 

It might be possible, and works for certain cases, that the Jordan 
isomorphism theorem should be replaced by Wigner-In\"on\"u or Saletan 
contraction of symmetry groups. For a discussion see 
\cite{SallerFinkelstein,Gilmore}. The direction is to proceed 
from a higher symmetry group to subgroups by performing a certain 
limit. Well known is the case:
\begin{eqnarray}
\begin{array}{ccccc}
\mbox{De Sitter} & & \mbox{Poincar\'e} & & \mbox{Galilei} \\
SO(3,2)        & & & & \\
               & \searrow & & & \\
               &        & ISO(3,1) & \\
               & \nearrow &  & \searrow & \\
(SO(4,1))      &        &     &  & G(3) \\
               &        &   & \nearrow & \\
               &        & (ISO(4)) & \\[1ex]
               & \frac{c}{R} \rightarrow \infty & & c \rightarrow \infty &
\end{array}
\end{eqnarray}
The direction in this type of contraction is from the simple group (also
providing simple equations and physical laws) to more complicated 
structures. While one has no translations in the De Sitter group, 
such transformations occur in the inhomogeneous Poincar\'e group. Breaking 
down once more the group, yields the Galilei group. But the dynamical 
equations become tremendous complicated due to `relativistic corrections'
and even wrong if the limit ($c \gg v$) is not valid.

Furthermore such contractions occur also in quantum physics. One can 
obtain the Heisenberg algebra from the following contraction scheme,
where $J_+$, $J_-$, $J_3$, $\xi=I$, generating $SL(n)$, is contracted:
\begin{eqnarray}
\left[\begin{array}{c}
h_+ \\ h_- \\ h_3 \\ I
\end{array}\right]
=
\left[\begin{array}{cccc}
\alpha & & & \\
 & \alpha & & \\
 & & 1 + \frac{\beta}{2\alpha} & \\
 & & & 1
\end{array}\right]
\quad
\left[\begin{array}{c}
J_+ \\ J_- \\ J_3 \\ \xi
\end{array}\right]
\end{eqnarray}
Identifying $h_3=N$, $h_+=a^\dagger$, $h_-=a$ and $I=I$, we obtain the 
usual oscillator algebra $[N,a^\dagger]_-=a^\dagger$, $[N,a]_-=-a$ and
$[a^\dagger,a]_-=I$. The same process takes place in the `contraction'
of the 5-dimensional projective Kaluza-Klein theory. In algebraic
terms, this is a process going from finer to coarser gradings in the
algebras by factoring out certain even sub-algebras. Hestenes 
\cite{Hes-linalg} after Weyl \cite{WeylSTM}
calls this a space-time split. A projective interpretation of the
Lorentz group is also possible \cite{Gschwind-RZG,Con-thesis}. 

\paragraph{Metrics from 6-vectors}

A recent program tries to involve so called 6-vectors to derive 
metric fields \cite{Harnett1,Harnett2,HOR,OH}. Starting point is 
electrodynamics in media. The linear relation assumed between 
$\vec{E}$, $\vec{D}$ and $\vec{H}$, $\vec{B}$ is then sufficient 
to induce a Minkowski metric. However, a 6-vector ($\vec{E},\vec{B})$
is a screw \cite{Ball}, and screws played an important role in the 
mechanics of rigid bodies in curved space \cite{Ziegler}. Following up 
this idea, on comes to two loose ends. Screws are spinors and screws are 
special linear complexes. The approach of \cite{HOR} employs thus line 
geometry, and with Felix Klein one says that `line geometry is metrical 
geometry' (on the Null quadric). The next paragraph will try to connect 
this again to projective geometry.

\paragraph{Null-invariance and Field equations}

A radical projective approach has to provide own arguments to find 
field equations which should not depend on metrical properties. Till
now we rejected the metric theory of Einstein-Hilbert in 5-dimensional
space, but we gave no alternative. But it is known, that 
$\five{R}_{\mu\nu}=0$ is needed to obtain the (almost) correct contraction
to usual 4-dimensional Einstein-Hilbert-Maxwell theory.

In the last paragraph we noted, that 6-vectors are special linear 
complexes, see also \cite{Gschwind-LK,Wil}. In addition, our examples of projective 
geometry have been of 3-dimensional projective space which is described 
by 4 homogenous coordinates. This seems to be not related to 5-dimensional
Kaluza--Klein theories. However, the space of lines is already four 
dimensional, and the space of linear complexes constitutes a 
5-dimensional projective space which can be described by 6-vectors. 
If one demands, that the incidence relations of the 5-dimensional 
projective space of linear complexes can be reduced to the space of 
3-dimensional projective space {\it including} the incidence 
relations, one obtains a null-invariance \cite{Stosz1,Stosz2}. The
null-invariance describes a quadric and a metrical surface. If the
infinite elements are moved, the quadric becomes dynamical and the 
metric itself a field. The null-invariance is exactly given by an 
equation like $\five{R}_{\mu\nu}=0$, which we propose to re-interpret 
in this way. However, a deep understanding of this process
will need a much better understanding of the different duality structures
in GC algebras including their Hopf algebra nature, which is intimately
linked to the incidence structure.

\section{Conclusion}

We gave a selection of arguments, others are available, which are quoted
to point out different aspects of projective reasoning and its connection
to different fields of physics. Adding them together yields a picture from
a puzzle if correctly assembled. 

We proposed to replace the Jordan isomorphy theorem by Wigner-In\"on\"u 
or Saletan contractions of Lie groups. Furthermore one should interpret 
the `vacuum' field equations of 5-dimensional Kaluza-Klein theory as 
the statement of null invariance in the 5-dimensional projective space
of linear complexes. This is a radical change in the usually adopted
interpretation of Kaluza-Klein theories. We start from projective 3-space,
no artificial higher dimensional `world dimensions' are introduced. But,
this restricts our approach to 5-dimensional Kaluza-Klein theory. Such
a reasoning is in accord with the approach of \cite{HOR,OH}.

This program is not only motivated by projective relativity theory, but 
also by the puzzling fact that quantum field theory has via its recently
observed Hopf algebraic structure a close relation to projective geometry 
too! Perhaps, one cannot unify relativity and quantum mechanics, but there
is already a glimpse of a projective framework capable to describe both
structure. This article is a first small step in this new direction.

\section*{Acknowledgement}

I would like to thank Prof. G. Lessner for fruitful discussions. 

\end{document}